\documentclass[a4paper,11pt]{article}
\usepackage{pos}
\usepackage[percent]{overpic}
\usepackage{xcolor}
\usepackage{transparent}
\usepackage{float} 
\usepackage{booktabs}
\usepackage[load-configurations = abbreviations]{siunitx}
\usepackage{graphicx}
\usepackage{subcaption}
\usepackage[numbers]{natbib}

\title{Fifteen years of Gamma-Ray Burst observations at Very High Energies with H.E.S.S.}

\author*[a]{Edna Ruiz-Velasco}
\author[b]{Cornelia Arcaro}
\author[c]{Mathieu de Bony de Lavergne}
\author[d]{Zhi-Qiu Huang}
\author[a]{David Sanchez}
\author[e]{Mohanraj Senniappan}
\author[g]{Stefan Wagner}
\author[f]{Sylvia Zhu}
\onbehalf{on behalf of the H.E.S.S. collaboration.}
\affiliation[a]{Laboratoire d’Annecy de Physique des Particules (LAPP), Université Savoie Mont Blanc, CNRS/IN2P3, 9 Chemin de Bellevue, 74940 Annecy, France.}
\affiliation[c]{Centre de Physique des Particules de Marseille (CPPM)f, Aix–Marseille Université, CNRS/IN2P3, 163 Avenue de Luminy – Case 902, 13288 Marseille cedex 9, France.}
\affiliation[b]{INFN Sezione di Padova and Department of Physics and Astronomy, University of Padova, Via Marzolo 8, 35131 Padova, Italy.}
\affiliation[d]{Scuola Internazionale Superiore di Studi Avanzati (SISSA), Via Bonomea 265, 34136 Trieste, Italy.}
\affiliation[e]{Department of Physics, Linnaeus University, SE-351 95 Växjö, Sweden.}
\affiliation[f]{Deutsches Elektronen-Synchrotron (DESY), Platanenallee 6, 15738 Zeuthen, Germany.}
\affiliation[g]{Landessternwarte (LSW), Zentrum für Astronomie der Universität Heidelberg, Königstuhl 12, 69117 Heidelberg, Germany.}

\emailAdd{hess-contact@mpi-hd.mpg.de}

\abstract{%
We present results from the High Energy Stereoscopic System (H.E.S.S.) follow-up observations of Gamma-ray Bursts (GRBs) between 2004 and 2019. We are focusing on non-detections and providing the most extensive set of very-high-energy (VHE, >100 GeV) upper limits to date. We use this catalogue to constrain the properties of VHE-detected GRBs and compare them to those detected at VHE. Our study finds that VHE-detected GRBs are not a distinct population but are instead associated with bright X-ray afterglows and low redshifts. In addition, we model the multi-wavelength emission of a few of the observed GRBs and discuss the results in the context of their obtained microphysical parameters. The results from this work help put current VHE observations into perspective and highlight the capabilities of next-generation instruments, in detecting fainter and more distant GRBs at VHE.}

\ConferenceLogo{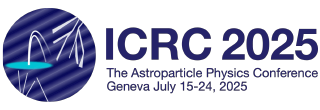}

\FullConference{39th International Cosmic Ray Conference (ICRC2025)\\
 15–24 July 2025\\
 Geneva, Switzerland\\}

\begin{document}
\maketitle

\tableofcontents

\section{Introduction}
Gamma‐Ray Bursts (GRBs) are the most luminous explosions in the Universe. Their prompt emission, lasting a few seconds, is mainly detected in the keV–MeV energy range. Their afterglows span all the electromagnetic spectrum up to TeV energies. Imaging Atmospheric Cherenkov Telescopes (IACTs) such as H.E.S.S. detect gamma rays above 100 GeV, providing input for studies of particle acceleration, magnetic field amplification, and attenuation due to the extragalactic background light (EBL)~\cite{franceschini} in GRBs. Since the beginning of H.E.S.S. operations, the H.E.S.S. collaboration has conducted systematic follow‐ups of GRB triggers to characterise the occurrence rate and properties of their TeV emission. Dedicated publications have covered the detailed analyses of two H.E.S.S. detections (GRB 180720B and GRB 190829A), and GRB~221009A (detected at VHEs by LHAASO) \cite{HESS_GRB180720B,HESS_GRB190829A,HESS_upper190114C}. Here, we present a results of the observation of GRBs with H.E.S.S. spanning 15 years, from 2004 to 2019, focusing on the observations leading to no detections. We highlight the statistical distinctions between detected and non-detected GRBs and derive constraints on the physical parameters required for VHE emission in three specific GRBs.

\section{H.E.S.S.\ Telescopes and GRB Follow‐Up}
The High Energy Stereoscopic System (H.E.S.S.) is located at 23$^\circ$16' S, 16$^\circ$30' E (Namibia, 1800 m a.s.l.) and consists of five telescopes: CT1–CT4 (12 m diameter, operational since 2004) and CT5 (28 m, operational since 2012). CT1–CT4 provide a $\sim$5$^\circ$ diameter field of view (FoV) and an energy threshold of $\sim$100 GeV at zenith; CT5 can lower this threshold to $\sim$50 GeV \cite{HESS_instrument2006}. The fast‐slewing speed of CT5 (repointing time $<100$ s to any part of the sky) allows early observations of GRB positions upon receipt of GCN notices from \textit{Swift}, \textit{Fermi}, and other satellites. Observations are conducted in wobble mode (offset $\sim0.5^\circ$) to allow for simultaneous background estimation.

GRB observations at H.E.S.S. are carried out as soon as possible, after receiving a GCN alert from instruments such as \textit{Swift}/BAT or \textit{Fermi}/GBM. When the alert is received, the telescopes slew to the reported coordinates, typically within 100–300 s of the satellite trigger. During these observations, a real‐time analysis (RTA) pipeline monitors the events to search for a detected VHE signal. If a signal appears at $\gtrsim3\sigma$, the observing crew is notified to adjust the observing strategy or extend the exposure. More details on the H.E.S.S. transients programme can be found in Ref.\cite{cornejoICRC}.

The observing criteria for GRBs have evolved through the years. Globally, the criteria applied during the period covering this work are as follows: (1) the zenith angle at the start of a potential observation must be $<60^\circ$ to keep the energy threshold as low as possible. (2) none or moderate moonlight are preferred (3) if a redshift measurement is available, preference is given to bursts with $z\lesssim0.5$ to mitigate EBL attenuation, allowing for observations with longer delays. (4) if the redshift is not known, decisions for observations are made based on prompt fluence and estimated visibility, such that bursts with high X-ray fluence or long duration are prioritised.

\section{Data Selection and Analysis Methodology}
\label{sec:analysis}

In this analysis, the GRBs observed from 2004 to 2019 were considered. This sample was extended by cross-matching all H.E.S.S. observations with public GRB catalogues from \emph{Swift}, \emph{Fermi}/GBM, INTEGRAL, MAXI, and HETE-2. This method added two \emph{Fermi}/GBM GRBs detected serendipitously during other H.E.S.S. campaigns. For each GRB, H.E.S.S. runs were grouped into analysis clusters based on timing and observing position: observations starting within 10 minutes of the GRB trigger formed individual clusters, while runs separated by more than four hours or having significant pointing changes were analysed separately to avoid analysing early GRB signal with late observations of likely pure background. Data quality selection was performed on all data, primarily requiring stable hardware and favourable weather conditions. The data reconstruction and gamma/hadron separation is done with the (HAP) ImPACT and (ParisAnalysis) Model++ methods~\cite{deNaurois2009}, applying cuts optimised for soft photon index sources \cite{HESS_instrument2006}. For the estimation of the significance of detection, the ring background method \cite{Berge07} was used together with the Li\&Ma statistics. In cases with no significant excess ($<5\sigma$), 95\% C.L.\ upper limits are computed above a fixed energy threshold using the method of \cite{Rolke2005}, assuming a photon index $\Gamma=2.5$ without EBL correction. Differential upper limits presented in the detailed study of three GRBs (Sec. 4.4) were independently corrected for EBL absorption. For GRB observations with big localisation uncertainties (like those of \emph{Fermi}/GBM follow-ups), upper-limit maps were produced using the same CL and spectral assumption. The results of upper-limit maps will not be discussed in this contribution. 

\section{Results: Upper limits, Population Comparisons, Specific GRBs}
\label{sec:population}

\subsection{Upper limits summary}

No new GRB detections were found in this H.E.S.S. analysis. In general, observations started within minutes to a few hours post-trigger, with delays ranging from about 2 minutes to several hours. Energy thresholds varied between approximately 90\,GeV and 700\,GeV, depending on observing conditions and analysis configuration. The integral flux upper limits at 95\% C.L, typically ranged between $1\times10^{-12}$ and a few times $10^{-11},\mathrm{cm^{-2}s^{-1}}$, highlighting the sensitivity achieved in these searches.

\subsection{Population comparisons}

We test whether VHE detection shares distinctive properties compared to non-detections, and we also assess whether our follow-up strategy reflects any observational bias beyond the selection of GRBs with known/small redshifts. To do so, we compare the prompt and afterglow properties of H.E.S.S.–observed GRBs, including GRB 180720B and GRB 190829A, to the broader samples from \emph{Swift}/XRT, and \emph{Swift}/BAT. To estimate selection effects, we first confirmed that the H.E.S.S. follow-up sample is statistically consistent with the full satellite catalogues. We then used a two-sample Kolmogorov–Smirnov tests (significant at p < 0.05) to compare the entire X-ray catalogues of GRBs, those observed by H.E.S.S., and the subset detected at VHEs (in this case, we consider comparisons with and without GRB 190829A as it is the only VHE-GRB candidate for being a low-luminosity GRB).

\subsubsection{Prompt-phase properties: \emph{Swift}/BAT sample}

\begin{figure}
  \centering
  \begin{overpic}[width=0.8\textwidth]{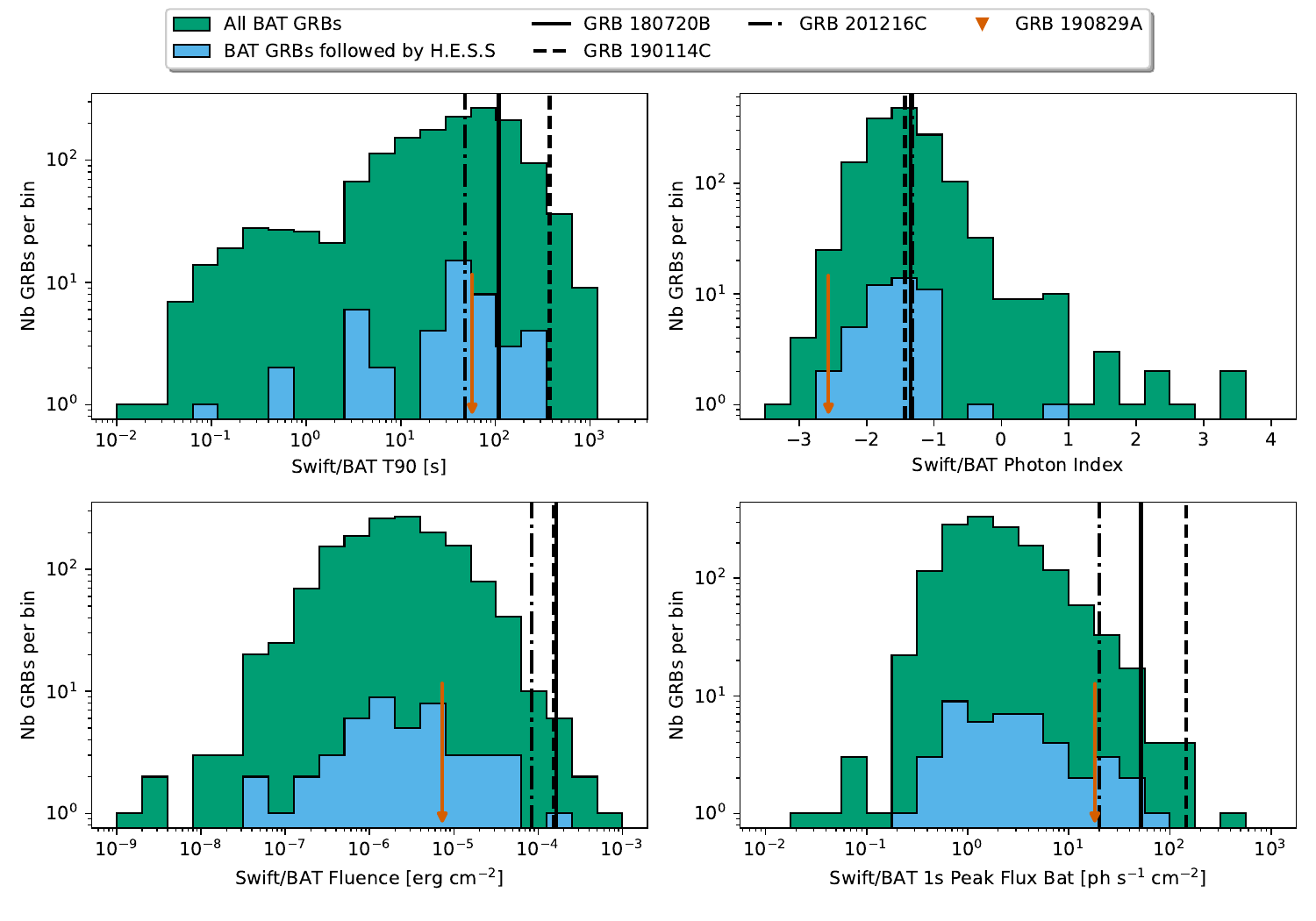}
    \put(50,30){%
    \makebox(0,0){%
      \rotatebox{-45}{%
        \transparent{0.4}%
        \textcolor{gray}{\fontsize{30}{20}\selectfont\bfseries PRELIMINARY}%
      }%
    }%
  }
\end{overpic}
  \caption{Distributions of \emph{Swift}/BAT parameters: whole BAT sample (green), H.E.S.S.–observed (blue), and VHE detections (black lines; GRB 190829A is indicated with an orange arrow). Top left: $T_{90}$. Top right: photon index. Bottom left: fluence. Bottom right: 1 s peak flux.}
  \label{fig:population_BAT}
\end{figure}

Table~\ref{tab:population_BAT} summarises the KS-test $p$-values for some of the \emph{Swift}/BAT prompt–phase parameters, showing no significant bias between H.E.S.S.–observed GRBs and the full BAT sample\footnote{GRB 221009A was outside BAT’s FoV}. The largest offset is $1.6\sigma$ in the 1\,s peak flux. Fig.~\ref{fig:population_BAT} compares $T_{90}$, photon index, fluence, and 1 s peak flux. In VHE detections, the fluence is the most distinct parameter ($p=2.5\times10^{-6}$, $4.7\sigma$; $2.9\sigma$ including GRB 190829A), with the 1\,s peak flux also showing a $3.9$–$4.5\sigma$ hint.

\begin{table}
    \centering
        \caption[]{$p$-value from the KS test for the \emph{Swift}/BAT observables
    \label{tab:population_BAT}}
    \resizebox{0.99\textwidth}{!}{

    \begin{tabular}{lccc}
    \toprule
    Parameter & $p$-value Observed & $p$-value Detected at VHE & $p$-value Detected at VHE (except GRB 190829A) \\
     & VS Whole population & VS Whole population & VS Whole population  \\
     \midrule
    $T_{90}$ & 0.16 ($1.4\sigma$)& 0.13 ($1.5\sigma$)& 0.25 ($1.2\sigma$)\\
    Fluence & 0.44 ($0.8\sigma$)& \SI{4.3e-3}{} ($2.9\sigma$)& \SI{2.5e-6}{} ($4.7\sigma$)\\
    1s Peak Flux & 0.10 ($1.6\sigma$)& \SI{6.2e-6}{} ($4.5\sigma$)& \SI{1.0e-4}{} ($3.9\sigma$)\\
    Spectral Index & 0.24 ($1.2\sigma$)& 0.58 ($0.6\sigma$)& 0.24 ($1.2\sigma$)\\
    \bottomrule
    \end{tabular}
 }
\end{table}

\subsubsection{Afterglow‐Phase Properties (\textit{Swift}/XRT Sample)}

\begin{figure}
\centering
\begin{overpic}[width=0.8\textwidth]{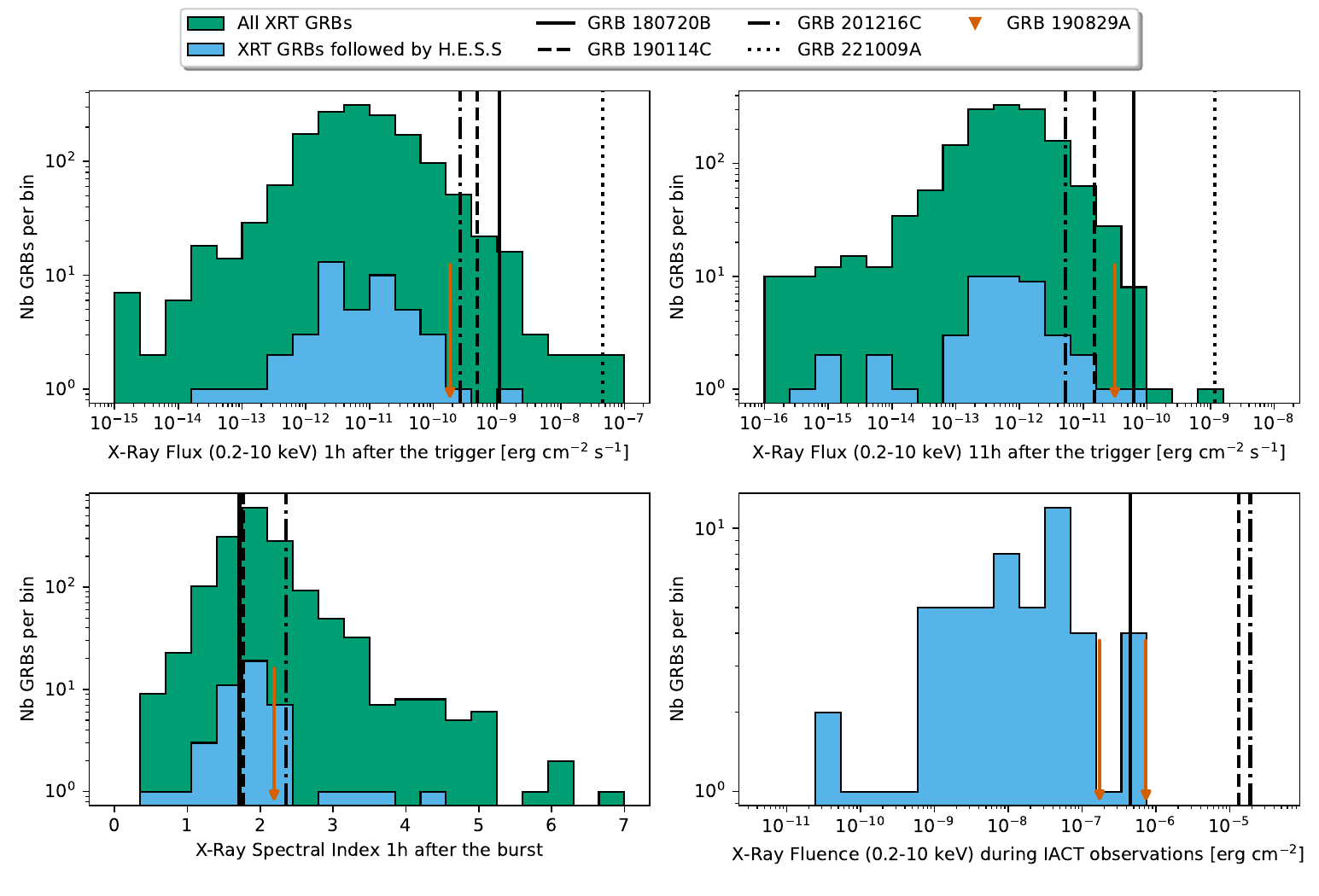}
  \put(50,30){%
    \makebox(0,0){%
      \rotatebox{-45}{%
        \transparent{0.4}%
        \textcolor{gray}{\fontsize{30}{20}\selectfont\bfseries PRELIMINARY}%
      }%
    }%
  }
\end{overpic}
\caption[]{Distribution of GRB properties measured by \emph{Swift}/XRT: green shows the full population, blue those observed by H.E.S.S. Black lines mark VHE-detected GRBs, with GRB190829A highlighted by orange arrows. Top left: $T_{90}$ distribution. Top right: total fluence. In the bottom right, the two orange arrows indicate the first and second nights of H.E.S.S. observations for GRB190829A. }
\label{fig:population_XRT}
\end{figure}

We used the live XRT catalogue to extract the flux level and spectral index at 1 h, and 11 h, and fluence during the H.E.S.S. window (extrapolating as needed, excluding GRB 221009A). The KS-test applied to each observable (Tab.~\ref{tab:population_XRT}) shows no bias between this H.E.S.S. and full XRT sample. Flux and fluence for VHE detections show $\gtrsim3\sigma$ deviations, indicating higher afterglow brightness. Fig.~\ref{fig:population_XRT} shows the distribution of these observables. 

\begin{table*}
    \centering
    \caption{$p$-value from the KS test for observables of \emph{Swift}/XRT.}
    \label{tab:population_XRT}
    \resizebox{0.99\textwidth}{!}{
        \begin{tabular}{lccc}
            \toprule
            Parameter & $p$-value Observed & $p$-value Detected at VHE & $p$-value Detected at VHE (except GRB 190829A) \\
            & vs whole population & vs whole population & vs whole population  \\
            \midrule
            Fluence during observations & N/A & \SI{6.5e-5}{} ($4.0\sigma$)& \SI{4.6e-5}{} ($4.1\sigma$)\\
            excluding GRB 221009A &  &  &  \\
            Flux at 1 h & 0.89 ($0.1\sigma$)& \SI{1.3e-6}{} ($4.0\sigma$)& \SI{6.6e-6}{} ($4.5\sigma$)\\
            Flux at 11 h & 0.92 ($0.1\sigma$)& \SI{9.7e-6}{} ($4.8\sigma$)& \SI{1.1e-4}{} ($3.8\sigma$)\\
            Spectral Index at 1 h & 0.48 ($0.7\sigma$)& 0.68 ($0.4\sigma$)& 0.31 ($1.0\sigma$)\\
            \bottomrule
        \end{tabular}
    }
\end{table*}

\subsection{Redshift and EBL Effects}
\begin{figure}
\centering
\begin{overpic}[width=0.45\textwidth]{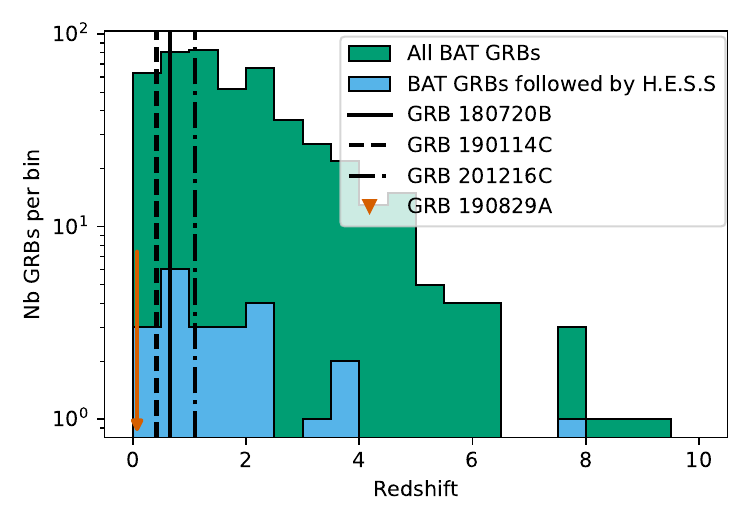}
  \put(60,30){%
    \makebox(0,0){%
      \rotatebox{0}{%
        \transparent{0.6}%
        \textcolor{gray}{\fontsize{15}{20}\selectfont\bfseries PRELIMINARY}%
      }%
    }%
  }
\end{overpic}

\caption[]{Redshift distribution for GRBs in the \emph{Swift}/BAT sample (green), those observed by H.E.S.S. (blue), and VHE detections (black lines). GRB~190829A is marked by the orange arrow.}
\label{fig:population_redshift}
\end{figure}

The detection of VHE photons in GRBs is limited by the redshift of the GRB due to the EBL absorption.
In this study, H.E.S.S. has observed 23 GRBs with a measured redshift, among these only 9 observations/detections occurred at $z<0.8$. As expected, this parameter is the only one showing a bias being introduced by our observation criteria. Compared to the full \emph{Swift}/BAT sample, GRBs observed or detected at VHE tend to have lower redshifts, showing a 2$\sigma$ difference (see Fig.~\ref{fig:population_redshift}). 

\subsection{Specific GRBs}

By applying a selection of known redshift, early H.E.S.S. observations (delay $<$1000\,s) and high \emph{Swift}/XRT fluence we select a subset of GRBs in this sample with well-constrained H.E.S.S.S. ULs. These GRBs were selected to perform a synchrotron self-Compton (SSC) modelling, placing the X-ray spectra and VHE ULs in a theoretical context. This selection yielded three cases: GRB100621A, GRB131030A, and GRB~161001A. \emph{Swift}/XRT spectra were fitted with absorbed power-law models and Galactic and local extinction, using PySPEC. The contemporaneous H.E.S.S. upper limits were EBL-corrected.

\begin{figure}[]
  \centering
  \setlength{\tabcolsep}{0pt}

  \begin{tabular}{ccc}
    \begin{overpic}[width=0.33\textwidth,  trim=7 8 8 7, clip]{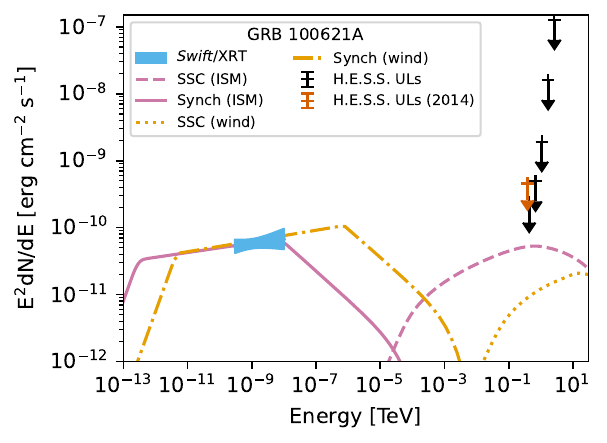}
      \put(60,30){%
        \makebox(0,0){%
          \rotatebox{-45}{%
            \transparent{0.3}%
            \textcolor{gray}{\fontsize{10}{20}\selectfont\bfseries PRELIMINARY}%
          }%
        }%
      }
    \end{overpic}
    &
    \begin{overpic}[width=0.33\textwidth, trim=7 8 8 7, clip]{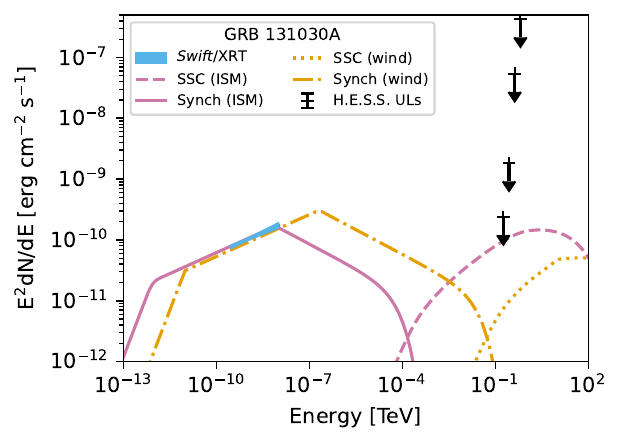}
      \put(60,30){%
        \makebox(0,0){%
          \rotatebox{-45}{%
            \transparent{0.3}%
            \textcolor{gray}{\fontsize{10}{20}\selectfont\bfseries PRELIMINARY}%
          }%
        }%
      }
    \end{overpic}
    &
    \begin{overpic}[width=0.33\textwidth, trim=7 8 8 7, clip]{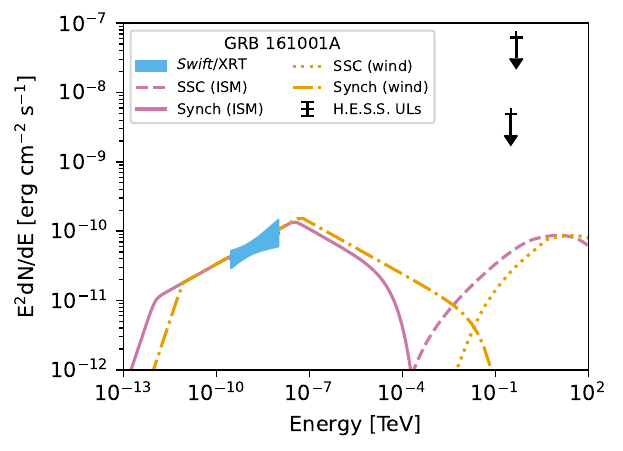}
      \put(60,30){%
        \makebox(0,0){%
          \rotatebox{-45}{%
            \transparent{0.3}%
            \textcolor{gray}{\fontsize{10}{20}\selectfont\bfseries PRELIMINARY}%
          }%
        }%
      }
    \end{overpic}
    \\
  \end{tabular}
  \caption{%
    \textbf{Left panel:} Spectral energy distribution of GRB~100621A. The \emph{Swift}/XRT spectrum is shown as a blue butterfly, H.E.S.S. ULs from this work in black, and the earlier H.E.S.S.\ UL from \cite{2014A&A...565A..16H} in orange. Pink solid (ISM) and yellow dot‐dashed (wind) curves show the synchrotron models; pink dashed (ISM) and yellow dotted (wind) curves show the SSC component.\\
    \textbf{Centre panel:} Spectral energy distribution of GRB~131030A. The \emph{Swift}/XRT butterfly is blue, H.E.S.S.\ ULs are black. Pink solid (ISM) and yellow dot‐dashed (wind) curves represent the synchrotron contribution; pink dashed (ISM) and yellow dotted (wind) curves are the SSC component.\\
    \textbf{Right panel:} Spectral energy distribution of GRB~161001A. The \emph{Swift}/XRT butterfly is blue, H.E.S.S.\ UL is black. Pink solid (ISM) and yellow dot‐dashed (wind) lines depict synchrotron; pink dashed (ISM) and yellow dotted (wind) lines depict SSC.%
  }
  \label{fig:three_grb_models}
\end{figure}

Emission was modelled using a standard single-zone SSC framework~\citep{1998ApJ...497L..17S,2001ApJ...548..787S,2022ApJ...925..182H}, with shock evolution following~\cite{1976PhFl...19.1130B}, and considering both constant-density (ISM) and wind-like environments. A representative set of parameters, required to reach the H.E.S.S. ULs as close as possible typically required low magnetic partition fractions ($\epsilon_B$ of $10^{-6}$–$10^{-5}$), electron partition fractions $\epsilon_e$ of 0.08–0.2, and ambient densities $n_0 \sim 2\times10^{-4}$ cm$^{-3}$ (ISM) or $A \sim 10^{33}$ cm$^{-1}$ (wind). The total shock energy required is $E_{\rm sh}$ was $3$–$6\times10^{55}$ erg. In all three GRBs, the X-ray spectra were consistent with synchrotron emission, while the H.E.S.S. upper limits were used to constrain the SSC component for both ISM and wind scenarios (see Fig.~\ref{fig:three_grb_models}).

\section{Discussion}

Between 2004 and 2019, H.E.S.S.\ followed up 46 GRBs (21 with known redshifts, $z=0.03$–6.3), detecting VHE emission from GRB180720B ($z=0.653$) and GRB190829A ($z=0.0785$)~\cite{HESS_GRB180720B,HESS_GRB190829A}. Our KS test indicates that VHE detections are associated with both high prompt peak flux and bright X-ray afterglows, supporting scenarios where dense seed photons boost SSC emission. The observed H.E.S.S. sample shows an expected bias to GRBs with low redshift; this choice is proved reasonable as only GRBs at low redshift are detected, consistent with EBL attenuation. The current H.E.S.S. sensitivity limits detection to nearby, high-energy bursts.  These results indicate that VHE-bright GRBs are rare, highlighting the importance of rapid response and focusing on bright, low-$z$ targets for future VHE searches.
Three well-observed GRBs were modelled with a single-zone SSC scenario, showing that the hard XRT spectra in combination with requiring to reach the H.E.S.S. upper limits with the SSC component pose a challenge to the explosion energies required. The absence of significant VHE signals in most bursts constrains the fraction of GRBs with favourable microphysics.


\begin{figure}[H]
\centering
\begin{subfigure}[t]{0.48\textwidth}
    \centering
    \begin{overpic}[width=\textwidth]{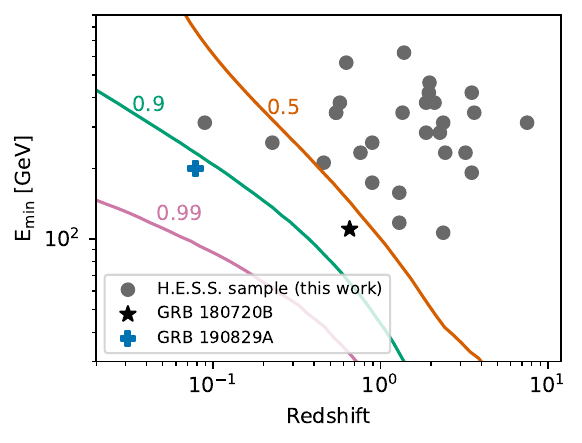}
    \put(55,45){%
        \makebox(0,0){%
          \rotatebox{-45}{%
            \transparent{0.3}%
            \textcolor{gray}{\fontsize{15}{20}\selectfont\bfseries PRELIMINARY}%
          }%
        }%
      }
    \end{overpic}
    \caption{Energy threshold and redshift for the H.E.S.S. sample. Curves show the energy at which EBL absorption reaches 0.99 (pink), 0.9 (turquoise), and 0.5 (orange). VHE-detected GRBs are marked.}
    \label{fig:ethr_ebl}
\end{subfigure}
\hfill
\begin{subfigure}[t]{0.48\textwidth}
    \centering
    \begin{overpic}[width=\textwidth]{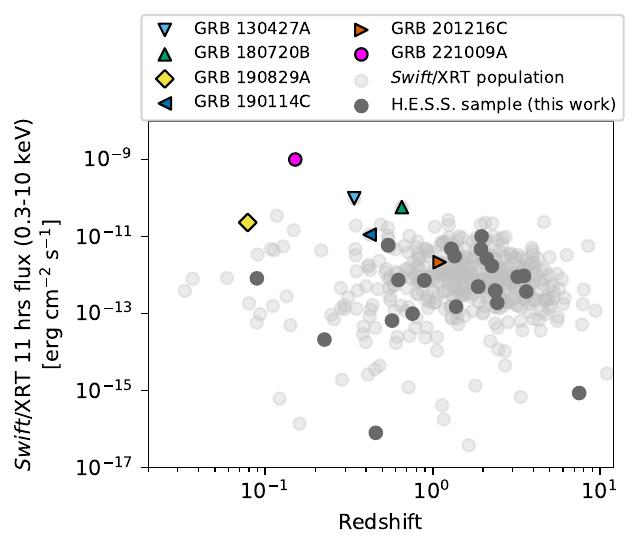}
        \put(60,40){%
        \makebox(0,0){%
          \rotatebox{-45}{%
            \transparent{0.3}%
            \textcolor{gray}{\fontsize{15}{20}\selectfont\bfseries PRELIMINARY}%
          }%
        }%
      }
    \end{overpic}
    
    \caption{ \emph{Swift}/BAT 11-hour flux vs. redshift. Light grey: all \emph{Swift} GRBs; dark grey: H.E.S.S. sample; coloured: VHE detections and GRB~130427A.}
    \label{fig:XraysRedshift}
\end{subfigure}
\caption{Left: EBL attenuation and energy threshold for the H.E.S.S. sample. Right: X-ray afterglow flux versus redshift for GRBs in the \emph{Swift} and H.E.S.S. samples.}
\label{fig:ebl_xrt_panels}
\end{figure}

 As shown in Fig.~\ref{fig:XraysRedshift} and Fig.~\ref{fig:ethr_ebl}, and supported by our population comparisons, VHE-detected GRBs stand out for their high X-ray afterglow fluxes and low redshifts, lying well within the detection horizon set by EBL attenuation and the H.E.S.S. energy threshold. The lack of rapid, low-zenith observations, but most importantly, the intrinsic delays in satellite alerts further limit prompt-phase coverage, making most H.E.S.S. GRB observations sensitive only to bright afterglows. Overall, current IACTs are effectively probing only the most favourable GRBs, while many potential VHE emitters remain beyond reach due to observational constraints and EBL effects.

\section{Conclusions and Outlook}
We analysed 48 GRBs observed by H.E.S.S. in the period of 2004 to 2019. Our population comparisons show that VHE detections are limited to GRBs with high prompt flux, bright X-ray afterglow, and low redshift. Through  modelling of the X-ray spectra and H.E.S.S. ULs for three specific GRBs, we found that extremely high explosion shock energies are required to produce an SSC component capable or reaching the H.E.S.S. ULs. 
The Cherenkov Telescope Array Observatory (CTAO), with a factor of 5–10 improvement in sensitivity and faster slewing ($\sim$20 s) will enable more detections~\cite{CTA_GRB2019}. Detections of more distant, less bright GRBs are needed to probe EBL models and jet microphysics with better detail. 
An upcoming journal publication will dive deeper into the findings of this analysis of 15 years of GRB observations with H.E.S.S.

\textbf{Ackwnoledgements see \url{https://hess.in2p3.fr/acknowledgements/}}

\begingroup
\setlength{\itemsep}{0pt}
\setlength{\parskip}{0pt}
\bibliographystyle{unsrt}
\bibliography{references}

\begin{thebibliography}{10}

\bibitem{franceschini}
A.~et~al. Franceschini.
\newblock {\em A\&A}, 487:837, 2008.

\bibitem{HESS_GRB180720B}
H.E.S.S. Collaboration and H.~et~al. Abdalla.
\newblock {\em Nature}, 575:464--467, 2019.

\bibitem{HESS_GRB190829A}
H.E.S.S. Collaboration and H.~et~al. Abdalla.
\newblock {\em Science}, 372(6546):1081--1085, 2021.

\bibitem{HESS_upper190114C}
F.~Aharonian et~al. (H.E.S.S.~Collaboration).
\newblock {\em ApJL}, 946:L27, 2023.

\bibitem{HESS_instrument2006}
F.~et~al. Aharonian.
\newblock {\em Astron. Astrophys.}, 457:899--915, 2006.

\bibitem{cornejoICRC}
B.~Cornejo et~al. (H.E.S.S.~Collaboration).
\newblock In these proceedings.
\newblock In {\em Proc. ICRC}, 2023.

\bibitem{deNaurois2009}
M.~et~al. de~Naurois.
\newblock {\em Astropart. Phys.}, 32:231--252, 2009.

\bibitem{Berge07}
D.~et~al. Berge.
\newblock {\em A\&A}, 466:1219--1229, 2007.

\bibitem{Rolke2005}
W.~A.~Rolke et~al.
\newblock {\em Nucl. Instrum. Methods Phys. Res. A}, 551:493--503, 2005.

\bibitem{1998ApJ...497L..17S}
R.~et~al. Sari.
\newblock {\em ApJ}, 497:L17, 1998.

\bibitem{2001ApJ...548..787S}
R.~et~al. Sari.
\newblock {\em ApJ}, 548:787, 2001.

\bibitem{2022ApJ...925..182H}
Z.-Q. et~al. Huang.
\newblock {\em ApJ}, 925:182, 2022.

\bibitem{1976PhFl...19.1130B}
R.~D. et~al. Blandford.
\newblock {\em Physics of Fluids}, 19:1130, 1976.

\bibitem{2014A&A...565A..16H}
H.E.S.S.~Collaboration et~al.
\newblock {\em A\&A}, 565:A16, 2014.

\bibitem{CTA_GRB2019}
S.~et~al. Inoue.
\newblock {\em Astropart. Phys.}, 43:252--275, 2013.

\end{thebibliography}

\endgroup
\end{document}